\documentclass[10pt, twocolumn]{extarticle}
\usepackage{ametsoc2col}
\usepackage{abstract}
\newcommand{\myabstract}{
\vspace{12pt}  When studying tropical cyclones using the $f$-plane, 
axisymmetric, gradient balanced model, there arises a second-order 
elliptic equation for the transverse circulation. Similarly, when 
studying zonally symmetric meridional circulations near the equator 
(the tropical Hadley cells) or the katabatically forced meridional 
circulation over Antarctica, there also arises a second order elliptic 
equation. These elliptic equations are usually derived in the pressure  
coordinate or the potential temperature coordinate, since the thermal 
wind equation has simple non-Jacobian forms in these two vertical 
coordinates. Because of the large variations in surface pressure that 
can occur in tropical cyclones and over the Antarctic ice sheet, there 
is interest in using other vertical coordinates, e.g., the height coordinate, 
the classical $\sigma$-coordinate, or some type of hybrid coordinate 
typically used in global numerical weather prediction or climate models. 
Because the thermal wind equation in these coordinates takes a Jacobian 
form, the derivation of the elliptic transverse circulation equation is 
not as simple. Here we present a method for deriving the elliptic transverse 
circulation equation in a generalized vertical coordinate, which allows 
for many particular vertical coordinates, such as height, pressure, log-pressure,  
potential temperature, classical $\sigma$, and most hybrid cases. Advantages 
and disadvantages of the various coordinates are discussed.   
\vspace{12pt}}

\begin{document}
%
%
\title{\textbf{\large{Elliptic Transverse Circulation Equations for Balanced Models \\
                      in a Generalized Vertical Coordinate}}}
%
%
\author{\textsc{Wayne H.\ Schubert}
				\thanks{\textit{Corresponding author address:} 
				Wayne H.\ Schubert,\newline Department of Atmospheric Science,
                Colorado State University,\newline Fort Collins, Colorado, USA. 
				\newline{E-mail: waynes@atmos.colostate.edu}}\\
\textit{\footnotesize{Department of Atmospheric Science,
Colorado State University, Fort Collins, Colorado, USA}}
\and 
\centerline{\textsc{Scott R. Fulton}}\\
\centerline{\textit{\footnotesize{Department of Mathematics,
Clarkson University, Potsdam, New York, USA}}}
\and 
\centerline{\textsc{Paul E. Ciesielski}}\\
\centerline{\textit{\footnotesize{Department of Atmospheric Science,
Colorado State University, Fort Collins, Colorado, USA}}}
}
%
\ifthenelse{\boolean{dc}}
{
\twocolumn[
\begin{@twocolumnfalse}
\amstitle

\begin{center}
\begin{minipage}{13.0cm}
\begin{abstract}
	\myabstract
	\newline
	\begin{center}
		\rule{38mm}{0.2mm}
	\end{center}
	\vspace{12pt}
\end{abstract}
\end{minipage}
\end{center}
\end{@twocolumnfalse}
]\saythanks
}
{
\amstitle
\begin{abstract}
\myabstract
\end{abstract}
\newpage
}

\section{Introduction}

     The purpose of the present paper is to derive the elliptic transverse 
circulation equation for an $f$-plane gradient balanced model of a tropical 
cyclone and the elliptic meridional circulation equation for zonally 
symmetric circulations on the spherical earth. 
The concept of these types of balanced dynamics was first proposed by 
\citet{eliassen51}. Since both these elliptic equations can be expressed in 
many different vertical coordinates, we first derive them in a generalized vertical 
coordinate $\eta$, and then consider five different commonly used choices 
of $\eta$. The paper is organized into two main parts. Part I discusses the 
elliptic transverse circulation equation for the $f$-plane, gradient 
balanced model of a tropical cyclone. The equation is first derived in 
section 3 using the generalized vertical coordinate $\eta$, with equation 
(\ref{eq3.11}) being the main result. Then, five 
particular choices of $\eta$ are discussed in sections 4--8. Part II 
discusses the elliptic equation for the meridional circulation 
on the spherical earth, with the goal of applications to the tropical 
Hadley cells and to the forced meridional circulation over Antarctica. 
Again, the equation is first derived in section 10 using the generalized 
vertical coordinate $\eta$, with equation (\ref{eq10.11}) being the main result.  
The five particular choices of $\eta$ are discussed in sections 11--15. 
A brief roadmap to the paper is given in Table 1.

\bigskip
\centerline{\bf Part I: Tropical Cyclones}

\section{The balanced vortex model in $(r,\eta,t)$}     

     For simplicity, the present analysis omits frictional effects. 
To simplify the primitive equation model to a balanced vortex model 
we assume that the azimuthal flow remains in a gradient balanced state, 
i.e., we discard the radial equation of motion and replace it with the 
gradient balance condition given below in (\ref{eq2.1}).  A sufficient 
condition for the validity of this assumption is that the diabatic 
forcing effects have slow enough time scales that significant, azimuthal 
mean inertia-gravity waves are not excited, i.e., $|Du/Dt|$ 
remains small compared to the pressure gradient and Coriolis/centrifugal terms. 
We shall describe this inviscid flow using a generalized vertical coordinate 
$\eta$, which is as yet an unspecified function of $(z,p,p_{_S},\theta)$, 
where $z$ is the height, $p$ the pressure, $p_{_S}$ the surface pressure, 
and $\theta$ the potential temperature.   
Under the balance condition, and using $\eta(z,p,p_{_S},\theta)$ as the 
vertical coordinate, the governing equations are  
\begin{equation}                           
        \left(f + \frac{v}{r}\right)v 
             =       \frac{\partial\Phi}{\partial r}
             + \theta\frac{\partial\Pi }{\partial r},        
\label{eq2.1}
\end{equation}
\begin{equation}                           
                  \frac{\partial v}{\partial t} 
      + \dot{\eta}\frac{\partial v}{\partial\eta} 
      + \left(f + \frac{\partial(rv)}{r\,\partial r}\right)u = 0,     
\label{eq2.2}
\end{equation}
\begin{equation}                          
      \frac{\partial\Phi}{\partial\eta} + \theta\frac{\partial\Pi}{\partial\eta} = 0,        
\label{eq2.3}
\end{equation}
\begin{equation}                          
          \frac{\partial m}{\partial t}
        + \frac{\partial (mru)}{r\,\partial r}
        + \frac{\partial (m \dot{\eta})}{\partial\eta} = 0,     
\label{eq2.4}
\end{equation}
\begin{equation}                          
                \frac{\partial\theta}{\partial t} 
    +         u \frac{\partial\theta}{\partial r} 
    + \dot{\eta}\frac{\partial\theta}{\partial\eta} = \dot{\theta},  
\label{eq2.5}
\end{equation}
where $f$ is the constant Coriolis parameter, $g$ the acceleration of gravity, 
$m=-(1/g)(\partial p/\partial\eta)$ the pseudo-density, $p$ the pressure, 
$\alpha=RT/p$ the specific volume, $\theta$ the potential temperature, $\Phi$ 
the geopotential, $\Pi=c_p(p/p_0)^\kappa$ the Exner function, $u$ the radial 
velocity component, $v$ the azimuthal velocity component, $\dot{\eta}$ the 
vertical velocity component, and where the partial derivatives $\partial/\partial t$ 
and $\partial/\partial r$ are understood to be at fixed $\eta$. 

     In the generalized vertical coordinate $\eta$, the radial pressure gradient 
force splits into two parts, as given on the right hand side of (\ref{eq2.1}).  
We have chosen to express the second part in terms of $\Pi$ and $\theta$, but 
$p$ and $\alpha$ could also be used since $\theta(\partial\Pi/\partial r)
=\theta(d\Pi/dp)(\partial p/\partial r)=\alpha(\partial p/\partial r)$. There are 
several vertical coordinates for which the radial pressure gradient force can be 
expressed as a single term. For example, the choice 
$\eta=z$ corresponds to the use of the height coordinate, in which case the 
radial pressure gradient force is $\alpha(\partial p/\partial r)_z$, since 
$(\partial\Phi/\partial r)_z=0$. Similarly, the choice $\eta=p$ corresponds 
to the use of the pressure coordinate, in which case the radial pressure gradient 
force is $(\partial\Phi/\partial r)_p$, since $(\partial\Pi/\partial r)_p=0$. 
Finally, the choice $\eta=\theta$ corresponds to the use of the potential 
temperature coordinate, in which case the two terms on the right hand side of 
(\ref{eq2.1}) combine to give $(\partial M/\partial r)_\theta$, where $M=\theta\Pi+\Phi$ 
is the Montgomery potential. 

    As we shall see in the following sections, it is the form of the radial pressure 
gradient force that determines the form of the thermal wind constraint and hence the 
mathematical intricacy in the derivation of the transverse circulation equation.  
For example, the simple forms of the radial pressure gradient force in the 
$p$-coordinate (or coordinates that are functions of $p$, such as $\ln p$) 
and the $\theta$-coordinate lead to simple forms of the thermal wind constraint 
and therefore straightforward derivations of the transverse circulation equation. 
This is in contrast to the $z$-coordinate, where the $\alpha$ coefficient in 
$\alpha(\partial p/\partial r)_z$ means that the thermal wind constraint involves 
a Jacobian, so the derivation of the transverse circulation equation is somewhat 
more complicated. With these observations in mind, the approach adopted here is 
to derive the transverse circulation equation in the generalized vertical coordinate 
(section 3), and then consider the special cases discussed in sections 4--8.

\section{Transverse circulation equation in $(r,\eta,t)$}     

\begin{table*}[b!t]                        
\centering
{\begin{tabular}{|c|c|c|c|c|}  
\hline\hline 
          Coordinate      &     Definition   &   Pressure Gradient & Thermal Wind &  Sections \\ 
\hline\hline 
           generalized    & $\eta(z,p,p_{_S},\theta)$                               
	                  & $\displaystyle{\left(\frac{\partial\Phi}{\partial r}\right)_\eta 
			    + \theta\left(\frac{\partial\Pi}{\partial r}\right)_\eta}$       
			  & $\displaystyle{\hat{f}\frac{\partial v}{\partial\eta}
			    =\frac{\partial(\Pi,\theta)}{\partial(r,\eta)}}$
			  &  3 and 10 \\[2.5ex]
            height        & $\eta=z$                                                
	                  & $\displaystyle{\frac{1}{\rho}\left(\frac{\partial p}{\partial r}\right)_z}$       
			  & $\displaystyle{\hat{f}\frac{\partial v}{\partial z}
			    =\frac{\partial(\Pi,\theta)}{\partial(r,z)}}$
			  &  4 and 11 \\[2.5ex]
        log-pressure      & $\eta=H\ln(p_0/p)\equiv\hat{z}$                        
	                  & $\displaystyle{\left(\frac{\partial\Phi}{\partial r}\right)_{\hat{z}}}$ 
			  & $\displaystyle{\hat{f}\frac{\partial v}{\partial\hat{z}}
			    =\frac{g}{T_0}\left(\frac{\partial T}{\partial r}\right)_{\hat{z}}}$
			  &  5 and 12 \\[2.5ex]
        pseudo-height     & $\eta=(c_p\theta_0/g)[1-(p/p_0)^\kappa]\equiv\tilde{z}$ 
	                  & $\displaystyle{\left(\frac{\partial\Phi}{\partial r}\right)_{\tilde{z}}}$   
			  & $\displaystyle{\hat{f}\frac{\partial v}{\partial\tilde{z}}
			    =\frac{g}{\theta_0}\left(\frac{\partial\theta}{\partial r}\right)_{\tilde{z}}}$ 
			  &  6 and 13 \\[2.5ex]
   potential temperature  & $\eta=\theta$                                           
                          & $\displaystyle{\left(\frac{\partial M}{\partial r}\right)_\theta}$        
			  & $\displaystyle{\hat{f}\frac{\partial v}{\partial\theta}
			    =\left(\frac{\partial\Pi}{\partial r}\right)_\theta}$
			  &  7 and 14 \\[2.5ex]
           sigma          & $\eta=1-(p/p_{_S})\equiv\sigma$                         
	                  & $\displaystyle{\left(\frac{\partial\Phi}{\partial r}\right)_\sigma
			    + \frac{(1-\sigma)}{\rho}\frac{\partial p_{_S}}{\partial r}}$ 
			  & $\displaystyle{\hat{f}\frac{\partial v}{\partial\sigma}
			    =\frac{\partial(\Pi,\theta)}{\partial(r,\sigma)}}$
			  &  8 and 15 \\[2.5ex]
\hline 
\end{tabular}}
\caption{The generalized vertical coordinate $\eta$ is a function of height $z$, 
pressure $p$, surface pressure $p_{_S}$, and potential temperature $\theta$. 
Five different vertical coordinates are discussed in sections 4--8 for the tropical 
cyclone problem and in sections 11--15 for the zonally symmetric meridional circulation 
problem. The pressure gradient 
force and the thermal wind equation (for the tropical cyclone case) in each coordinate are 
given in the third and fourth columns. Note that the thermal wind equation has a Jacobian 
form for the $z$ and $\sigma$ coordinates, but a non-Jacobian form for the $\hat{z}$, $\tilde{z}$, 
and $\theta$ coordinates. The derivation of the elliptic equation for the transverse circulation 
is more intricate in vertical coordinates for which the thermal wind equation takes a Jacobian 
form.}
\end{table*} 

     To derive the transverse circulation equation associated with 
(\ref{eq2.1})--(\ref{eq2.5}), first consider the mass continuity equation 
(\ref{eq2.4}), which can be written in the form   
\begin{equation}                              
       \frac{\partial(mru)}{r\,\partial r} 
     + \frac{\partial(m\hat{\dot{\eta}})}{\partial\eta} = 0,  
\label{eq3.1}
\end{equation}
where 
\begin{equation}                                    
       m\hat{\dot{\eta}} = m\dot{\eta} - \frac{1}{g}\frac{\partial p}{\partial t}.  
\label{eq3.2}
\end{equation}
Using this form of the mass conservation principle we define a streamfunction 
$\psi$ such that 
\begin{equation}                                    
              m u       = -\frac{\partial \psi}{\partial\eta},  \qquad  
      m\hat{\dot{\eta}} =  \frac{\partial(r\psi)}{r\,\partial r}.   
\label{eq3.3}
\end{equation}
In the following sections we consider only those cases for which the vertical coordinate 
$\eta$ increases in the upward direction, so that $m=-(1/g)(\partial p/\partial\eta)>0$.  

     The thermal wind equation, derived from (\ref{eq2.1}) and (\ref{eq2.3}), is 
\begin{equation}                                  
        \hat{f}\frac{\partial v}{\partial\eta} 
      = \frac{\partial(\Pi,\theta)}{\partial(r,\eta)},  
\label{eq3.4}
\end{equation}
where $\hat{f}=f+2v/r$. Taking $\partial/\partial t$ of (\ref{eq3.4}) we obtain 
\begin{equation}                                 
      \frac{\partial}{\partial\eta}\left(\hat{f}\frac{\partial v}{\partial t  }\right) 
    = \frac{\partial}{\partial t}
        \left(\frac{\partial(\Pi,\theta)}{\partial(r,\eta)}\right),   
\label{eq3.5}
\end{equation}
where we have made use of the identity 
$(\partial/\partial t)  [\hat{f}(\partial v/\partial\eta)]
=(\partial/\partial\eta)[\hat{f}(\partial v/\partial t  )]$ to express the 
left hand side of (\ref{eq3.5}) in a form that has $(\partial v/\partial t)$ 
inside the $(\partial/\partial\eta)$ operator. This sets the stage for the elimination 
of $(\partial v/\partial t)$ through the use of (\ref{eq2.2}). Similarly, it 
is convenient to rewrite the right hand side of (\ref{eq3.5}) in a form that 
has local time derivatives inside the $(\partial/\partial r)$ and 
$(\partial/\partial\eta)$ operators. This can be accomplished through the 
use of the Jacobi identity 
\begin{equation}                                     
    \frac{\partial}{\partial r  }\left(\frac{\partial(\Pi,\theta)}{\partial(\eta, t  )}\right)
  + \frac{\partial}{\partial\eta}\left(\frac{\partial(\Pi,\theta)}{\partial(  t , r  )}\right)
  + \frac{\partial}{\partial t  }\left(\frac{\partial(\Pi,\theta)}{\partial(  r ,\eta)}\right) = 0,   
\label{eq3.6}
\end{equation}
which allows (\ref{eq3.5}) to be written in the form 
\begin{equation}                                     
    \frac{\partial}{\partial r  }\left(\frac{\partial(\Pi,\theta)}{ \partial(\eta,t)}\right)
  + \frac{\partial}{\partial\eta}\left(\hat{f}\frac{\partial v}{\partial t}
                                      + \frac{\partial(\Pi,\theta)}{\partial(t,r)}\right) = 0.    
\label{eq3.7}
\end{equation}
This is the constraint that must be satisfied by the local time derivatives of the 
azimuthal wind field and the mass field. This constraint will lead directly to the elliptic 
equation for $\psi$, after we have derived equations for the two quantities in the large 
parentheses of (\ref{eq3.7}). The equation for the first of these is derived by 
multiplying (\ref{eq2.5}) by $(\partial\Pi/\partial\eta)$, while the equation for 
the second is derived by adding $\hat{f}$ times (\ref{eq2.2}) 
to $-(\partial\Pi/\partial r)$ times (\ref{eq2.5}). The results are 
\begin{equation}                                   
         \frac{\partial(\Pi,\theta)}{\partial(\eta,t)}
     = Am\hat{\dot{\eta}} - Bmu + \frac{\partial\Pi}{\partial\eta}\dot{\theta},   
\label{eq3.8}
\end{equation}
\begin{equation}                                   
     \hat{f}\frac{\partial v}{\partial t}
    + \frac{\partial(\Pi,\theta)}{\partial(t,r)}
    = Bm\hat{\dot{\eta}} - Cmu - \frac{\partial\Pi}{\partial r}\dot{\theta},    
\label{eq3.9}
\end{equation}
where  
\begin{equation}                                   
  \begin{split}
      A &= \frac{g\alpha}{\theta}\frac{\partial\theta}{\partial\eta},    \qquad 
      B  =-\frac{g\alpha}{\theta}\frac{\partial\theta}{\partial r},   \\
      C &= \frac{\hat{f}}{m}\left(f + \frac{\partial(rv)}{r\,\partial r}\right) 
         - \frac{1}{m}\frac{\partial\Pi}{\partial r}\frac{\partial\theta}{\partial r}.  
  \end{split}
\label{eq3.10}
\end{equation}
Using (\ref{eq3.8}) and (\ref{eq3.9}) in (\ref{eq3.7}) we obtain the meridional 
circulation equation 
\begin{equation}                                           
 \begin{split}
    &  \frac{\partial}{\partial r}
         \left( A \frac{\partial(r\psi)}{r\,\partial r} 
              + B \frac{\partial  \psi}{\partial\eta}\right)    \\
    & \quad 
     + \frac{\partial}{\partial\eta}
         \left( B \frac{\partial(r\psi)}{r\,\partial r} 
              + C \frac{\partial  \psi}{\partial\eta}\right)      
     = \frac{\partial(\Pi,\dot{\theta})}{\partial(r,\eta)}. 
 \end{split}   
\label{eq3.11}
\end{equation}
Note that $ AC-B^2=(g\alpha/\theta)\hat{f}P$, where 
\begin{equation}                                         
      P = \frac{1}{m}\left[-\frac{\partial v}{\partial\eta}\frac{\partial\theta}{\partial r}    
        + \left(f + \frac{\partial(rv)}{r\,\partial r}\right)\frac{\partial\theta}{\partial\eta}\right]    
\label{eq3.12}     
\end{equation}
is the potential vorticity. The partial differential equation (\ref{eq3.11}) is 
elliptic if $\hat{f}P>0$.

     In the next five sections we consider the special cases of the height coordinate, 
the log-pressure coordinate, the pseudo-height coordinate,  
the isentropic coordinate, and a particular form of the sigma 
coordinate, as summarized in Table 1.

\section{Transverse circulation equation in the height coordinate}  

     As the first special case, consider the height coordinate, $\eta=z$. 
In the height coordinate the pseudodensity equals the true density, i.e., 
$m=-(1/g)(\partial p/\partial z)=\rho$. The governing equations 
(\ref{eq2.1})--(\ref{eq2.5}) take the form   
\begin{equation}                           
     \left(f + \frac{v}{r}\right)v = \theta\frac{\partial\Pi}{\partial r},        
\label{eq4.1}
\end{equation}
\begin{equation}                           
         \frac{\partial v}{\partial t} 
      + w\frac{\partial v}{\partial z} 
      + \left(f + \frac{\partial(rv)}{r\,\partial r}\right)u = 0,     
\label{eq4.2}
\end{equation}
\begin{equation}                          
      \theta\frac{\partial\Pi}{\partial z} = -g,        
\label{eq4.3}
\end{equation}
\begin{equation}                          
          \frac{\partial\rho}{\partial t}
        + \frac{\partial (\rho ru)}{r\,\partial r}
        + \frac{\partial (\rho w)}{\partial z} = 0,     
\label{eq4.4}
\end{equation}
\begin{equation}                          
            \frac{\partial\theta}{\partial t} 
        + u \frac{\partial\theta}{\partial r} 
        + w \frac{\partial\theta}{\partial z} = \dot{\theta},  
\label{eq4.5}
\end{equation}
where in this section the partial derivatives $(\partial/\partial t)$ 
and $(\partial/\partial r)$ are taken at fixed $z$. The thermal wind 
equation (\ref{eq3.4}) becomes  
\begin{equation}                                  
        \hat{f}\frac{\partial v}{\partial z} 
      = \frac{\partial(\Pi,\theta)}{\partial(r,z)}.  
\label{eq4.6}
\end{equation}
The transverse circulation equation (\ref{eq3.11}) becomes  
\begin{equation}                              
  \begin{split}
    & \frac{\partial}{\partial r}
       \left( A \frac{\partial(r\psi)}{r\,\partial r} 
            + B \frac{\partial  \psi}{\partial z}\right)     \\
    & \qquad 
    + \frac{\partial}{\partial z}
       \left( B \frac{\partial(r\psi)}{r\,\partial r} 
            + C \frac{\partial  \psi}{\partial z}\right)
    = \frac{\partial(\Pi,\dot{\theta})}{\partial(r,z)},    
  \end{split}
\label{eq4.7}
\end{equation}
where, using the hydrostatic equation (\ref{eq4.3}), the coefficients 
(\ref{eq3.10}) become   
\begin{equation}                              
  \begin{split}
    \rho A &= \frac{g}{\theta}\frac{\partial\theta}{\partial z},   \qquad 
    \rho B  =-\frac{g}{\theta}\frac{\partial\theta}{\partial r},        \\
    \rho C &= \hat{f}\left(f + \frac{\partial(rv)}{r\,\partial r}\right)
            - \frac{\partial\Pi   }{\partial r}
              \frac{\partial\theta}{\partial r}.    
  \end{split}
\label{eq4.8}
\end{equation}
The mass flux (\ref{eq3.3}) becomes  
\begin{equation}                             
      \rho u = -\frac{\partial \psi}{\partial z},  \qquad  
      \rho w - \frac{1}{g}\frac{\partial p}{\partial t} =  \frac{\partial(r\psi)}{r\,\partial r},   
\label{eq4.9}
\end{equation}
and that $ AC-B^2=(g\alpha/\theta)\hat{f}P$, where 
\begin{equation}                             
    P = \frac{1}{\rho}\left[-\frac{\partial v}{\partial z}\frac{\partial\theta}{\partial r}    
      + \left(f + \frac{\partial(rv)}{r\,\partial r}\right)\frac{\partial\theta}{\partial z}\right].   
\label{eq4.10}
\end{equation} 
\citet{pendergrass09} and \citet{willoughby09} have presented an interesting analysis 
of the diabatically induced transverse circulation in tropical cyclones for both 
quasi-steady forcing and periodic forcing. Their formulation uses the height coordinate, 
which has the advantage that the lower boundary is a coordinate surface. Their mathematical 
analysis is generally similar to that presented here, although their derivation of the 
transverse circulation equation begins with the anelastic version of (\ref{eq4.4}).


\section{Transverse circulation equation in the log-pressure coordinate}   

    As the second special case, consider the the log-pressure coordinate, 
$\eta=\hat{z}=H\ln(p_0/p)$, where the scale height $H=RT_0/g$, the 
reference temperature $T_0$, and the reference pressure $p_0$ are constants.  
In this coordinate the pseudo-density becomes 
$m=\rho_0 e^{-\hat{z}/H} \equiv \hat{\rho}(\hat{z})$, where the constant 
reference density is $\rho_0=p_0/RT_0$.  The governing equations 
(\ref{eq2.1})--(\ref{eq2.5}) take the form    
\begin{equation}                           
        \left(f + \frac{v}{r}\right)v = \frac{\partial\Phi}{\partial r},        
\label{eq5.1}
\end{equation}
\begin{equation}                           
                \frac{\partial v}{\partial t} 
      + \hat{w} \frac{\partial v}{\partial\hat{z}} 
      + \left(f + \frac{\partial(rv)}{r\,\partial r}\right)u = 0,     
\label{eq5.2}
\end{equation}
\begin{equation}                          
      \frac{\partial\Phi}{\partial\hat{z}} = \frac{g}{T_0}T,        
\label{eq5.3}
\end{equation}
\begin{equation}                          
          \frac{\partial (\hat{\rho} ru)}{r\,\partial r}
        + \frac{\partial (\hat{\rho} \hat{w})}{\partial\hat{z}} = 0,     
\label{eq5.4}
\end{equation}
\begin{equation}                          
             \frac{\partial\theta}{\partial t} 
    +      u \frac{\partial\theta}{\partial r} 
    + \hat{w}\frac{\partial\theta}{\partial\hat{z}} = \dot{\theta},  
\label{eq5.5}
\end{equation}
where in this section the partial derivatives $(\partial/\partial t)$ 
and $(\partial/\partial r)$ are taken at fixed $\hat{z}$.
Since the pseudo-density $\hat{\rho}$ is a function of $\hat{z}$ only, 
the time derivative term in the continuity equation vanishes. 

    In contrast to (\ref{eq4.6}), the thermal wind equation (\ref{eq3.4})  
simplifies to the non-Jacobian form         
\begin{equation}                                  
        \hat{f}\frac{\partial v}{\partial\hat{z}} 
      = \frac{g}{T_0}\frac{\partial T}{\partial r}.  
\label{eq5.6}
\end{equation}
The mass flux (\ref{eq3.3}) becomes  
\begin{equation}                                    
           \hat{\rho} u  = -\frac{\partial \psi}{\partial\hat{z}},  \qquad  
       \hat{\rho}\hat{w} =  \frac{\partial(r\psi)}{r\,\partial r}.   
\label{eq5.7}
\end{equation}    
The last term in the second line of (\ref{eq3.10}) vanishes and 
the transverse circulation equation (\ref{eq3.11}) becomes  
\begin{equation}                                       
  \begin{split}
    & \frac{\partial}{\partial r}
       \left( A \frac{\partial(r\psi)}{r\,\partial r} 
            + B \frac{\partial  \psi }{ \partial\hat{z}}\right)  \\
    & \qquad 
     + \frac{\partial}{\partial\hat{z}}
       \left( B \frac{\partial(r\psi)}{r\,\partial r} 
            + C \frac{\partial  \psi }{\partial\hat{z}}\right)
    = \frac{T}{\theta}\frac{g}{T_0}\frac{\partial\dot{\theta}}{\partial r},    
  \end{split}
\label{eq5.8}
\end{equation}
where, using the hydrostatic equation (\ref{eq5.3}), the coefficients 
(\ref{eq3.10}) become  
\begin{equation}                                      
  \begin{split}
     \hat{\rho}A &= \frac{g}{T_0}\frac{\partial\theta}{\partial\hat{z}},   \quad 
     \hat{\rho}B  =-\frac{g}{T_0}\frac{\partial T}{\partial r}
                  =-\hat{f} \frac{\partial v}{\partial\hat{z}},  \\ 
     \hat{\rho}C &= \hat{f} \left(f + \frac{\partial(rv)}{r\,\partial r}\right).   
  \end{split}
\label{eq5.9}
\end{equation}
Note that $AC - B^2 = (g\alpha/\theta)\hat{f}P$, where
\begin{equation}                                        
     P = \frac{1}{\hat{\rho}}
         \left[-\frac{\partial v}{\partial\hat{z}}\frac{\partial\theta}{\partial r}    
              + \left(f + \frac{\partial(rv)}{r\,\partial r}\right)
	                  \frac{\partial\theta}{\partial\hat{z}}\right]    
\label{eq5.10}
\end{equation}
is the potential vorticity.
Examples of the use of the log-pressure coordinate are \citet{schubert07}, 
who studied the distribution of subsidence in the hurricane eye, and  
\citet{vigh09} and \cite{musgrave12}, who studied the rapid development 
of the tropical cyclone warm core. An obvious disadvantage of using $\hat{z}$ 
as the vertical coordinate in numerical modeling (although not so much in 
conceptual models) is that the lower boundary for (\ref{eq5.8}) is not 
generally a coordinate surface. 

\section{Transverse circulation equation in the pseudo-height coordinate}  

     Now consider the special case of the pseudo-height coordinate, $\eta=\tilde{z}$, 
where 
\begin{equation}                                  
       \tilde{z} = \frac{c_p\theta_0}{g} \left[1 - \left(\frac{p}{p_0}\right)^\kappa\right]. 
\label{eq6.1}
\end{equation}	    
This coordinate has been widely used in the study of frontogenesis, e.g., see 
\citet{hoskins+bretherton72}. Since $\tilde{z}=(\theta_0/g)(c_p-\Pi)$, we can 
regard $\tilde{z}$ as essentially an Exner function vertical coordinate, but 
scaled to have the unit of length and to increase in the upward direction.
In this coordinate the pseudo-density is given by 
\begin{equation}                                 
    m = \tilde{\rho}(\tilde{z}) = \frac{p_0}{R\theta_0}
                              \left(1 - \frac{g\tilde{z}}{c_p \theta_0}\right)^{(1-\kappa)/\kappa}.     
\label{eq6.2}
\end{equation}

     The governing equations (\ref{eq2.1})--(\ref{eq2.5}) take the form    
\begin{equation}                           
        \left(f + \frac{v}{r}\right)v = \frac{\partial\Phi}{\partial r},        
\label{eq6.2}
\end{equation}
\begin{equation}                           
                  \frac{\partial v}{\partial t} 
      + \tilde{w} \frac{\partial v}{\partial\tilde{z}} 
      + \left(f + \frac{\partial(rv)}{r\,\partial r}\right)u = 0,     
\label{eq6.3}
\end{equation}
\begin{equation}                          
      \frac{\partial\Phi}{\partial\tilde{z}} = \frac{g}{\theta_0}\theta,        
\label{eq6.4}
\end{equation}
\begin{equation}                          
          \frac{\partial (\tilde{\rho} ru)}{r\,\partial r}
        + \frac{\partial (\tilde{\rho} \tilde{w})}{\partial\tilde{z}} = 0,     
\label{eq6.5}
\end{equation}
\begin{equation}                          
               \frac{\partial\theta}{\partial t} 
    +        u \frac{\partial\theta}{\partial r} 
    + \tilde{w}\frac{\partial\theta}{\partial\tilde{z}} = \dot{\theta},  
\label{eq6.6}
\end{equation}
and the thermal wind equation (\ref{eq3.4}) becomes  
\begin{equation}                                  
        \hat{f}\frac{\partial v}{\partial\tilde{z}} 
      = \frac{g}{\theta_0}\frac{\partial\theta}{\partial r}.  
\label{eq6.7}
\end{equation}
The mass flux (\ref{eq3.3}) becomes  
\begin{equation}                                    
       \tilde{\rho}   u      = -\frac{\partial \psi}{\partial\tilde{z}},  \qquad  
       \tilde{\rho}\tilde{w} =  \frac{\partial(r\psi)}{r\,\partial r}.   
\label{eq6.8}
\end{equation}
The transverse circulation equation (\ref{eq3.11}) becomes  
\begin{equation}                                
  \begin{split}
    & \frac{\partial}{\partial r}
       \left( A \frac{\partial(r\psi)}{r\,\partial r} 
            + B \frac{\partial  \psi}{\partial\tilde{z}}\right)     \\
    & \quad 
    + \frac{\partial}{\partial\tilde{z}}
       \left( B \frac{\partial(r\psi)}{r\,\partial r} 
            + C \frac{\partial  \psi}{\partial\tilde{z}}\right)
    = \frac{g}{\theta_0}\frac{\partial\dot{\theta}}{\partial r},    
  \end{split}
\label{eq6.9}
\end{equation}
where, using the hydrostatic equation (\ref{eq6.3}), the coefficients 
(\ref{eq3.10}) become    
\begin{equation}                              
  \begin{split}  
    \tilde{\rho}A &= \frac{g}{\theta_0}\frac{\partial\theta}{\partial\tilde{z}},   \qquad 
    \tilde{\rho}B  =-\frac{g}{\theta_0}\frac{\partial\theta}{\partial r}
                   =-\hat{f}\frac{\partial v}{\partial\tilde{z}},        \\
    \tilde{\rho}C &= \hat{f}\left(f + \frac{\partial(rv)}{r\,\partial r}\right).        
  \end{split}
\label{eq6.10}
\end{equation}
Note that 
\begin{equation}                               
      \tilde{\rho}   u      = -\frac{\partial \psi}{\partial\tilde{z}},  \qquad  
      \tilde{\rho}\tilde{w} =  \frac{\partial(r\psi)}{r\,\partial r},  
\label{eq6.11}
\end{equation}
and $AC-B^2=(\alpha/\theta)\hat{f}P$, where 
\begin{equation}                               
    P = \frac{1}{\tilde{\rho}}\left[-\frac{\partial v}{\partial\tilde{z}}\frac{\partial\theta}{\partial r}    
      + \left(f + \frac{\partial(rv)}{r\,\partial r}\right)\frac{\partial\theta}{\partial\tilde{z}}\right].   
\label{eq6.12}
\end{equation}  
Examples of the use of the pseudo-height coordinate $\tilde{z}$ can be found in 
\citet{schubert+hack82} and \citet{hack+schubert86}, who studied the nonlinear response 
of atmospheric vortices to heating by organized cumulus convection.

\section{Transverse circulation equation in the isentropic coordinate}  

     Now consider the special case of the isentropic coordinate, $\eta=\theta$. In 
the isentropic coordinate the pseudo-density becomes $m=-(1/g)(\partial p/\partial\theta)$. 
The governing equations (\ref{eq2.1})--(\ref{eq2.5}) take the form    
\begin{equation}                           
        \left(f + \frac{v}{r}\right)v = \frac{\partial M}{\partial r},        
\label{eq7.1}
\end{equation}
\begin{equation}                           
                  \frac{\partial v}{\partial t} 
    + \dot{\theta}\frac{\partial v}{\partial\theta} 
    + \left(f + \frac{\partial(rv)}{r\,\partial r}\right)u = 0,     
\label{eq7.2}
\end{equation}
\begin{equation}                          
      \frac{\partial M}{\partial\theta} = \Pi,        
\label{eq7.3}
\end{equation}
\begin{equation}                          
          \frac{\partial m}{\partial t}
        + \frac{\partial (mru)}{r\,\partial r}
        + \frac{\partial (m\dot{\theta})}{\partial\theta} = 0,      
\label{eq7.4}
\end{equation}
where in this section the partial derivatives $(\partial/\partial t)$ 
and $(\partial/\partial r)$ are taken at fixed $\theta$.

    The thermal wind equation (\ref{eq3.4}) takes the non-Jacobian form   
\begin{equation}                                  
     \hat{f}\frac{\partial v}{\partial\theta} = \frac{\partial\Pi}{\partial r}.  
\label{eq7.5}
\end{equation}
The mass flux (\ref{eq3.3}) becomes 
\begin{equation}                                    
    mu      = -\frac{\partial \psi}{\partial\theta},  \qquad  
    m\dot{\theta} - \frac{1}{g}\frac{\partial p}{\partial t} = \frac{\partial(r\psi)}{r\,\partial r}.   
\label{eq7.6}
\end{equation}
Note that $A=g\alpha/\theta$, $B=0$, and the last term in the second line of (\ref{eq3.10}) 
vanishes so that $C=\hat{f}P$. The transverse circulation equation takes the simple form   
\begin{equation}                                    
      \frac{\partial}{\partial r}
       \left( \frac{g\alpha}{\theta} \frac{\partial(r\psi)}{r\,\partial r}\right) 
    + \frac{\partial}{\partial\theta}
       \left( \hat{f} P \frac{\partial\psi}{\partial\theta}\right)
    = \frac{\partial(\Pi,\dot{\theta})}{\partial(r,\theta)},       
\label{eq7.7}
\end{equation}
where  
\begin{equation}                                   
      P = \left(f + \frac{\partial(rv)}{r\,\partial r}\right)
          \left(-\frac{1}{g}\frac{\partial p}{\partial\theta}\right)^{-1}   
\label{eq7.8}
\end{equation}
is the potential vorticity. 

     Although the transverse circulation equation (\ref{eq7.7}) and 
the potential vorticity (\ref{eq7.8}) have compact forms in the 
$\theta$-coordinate, a disadvantage of using $\theta$ as the vertical 
coordinate is that the lower boundary of the atmosphere is generally 
not a $\theta$-surface. However, in $\theta$-coordinate modeling this 
disadvantage is sometimes overcome through the use of a massless layer, 
which effectively makes the lower boundary a $\theta$-surface. \citet{fulton91} 
have used this massless layer approach to study surface frontogenesis 
in isentropic coordinates. In the context of tropical cyclone dynamics, 
\citet{hendricks+schubert10} have used a generalized version of (\ref{eq7.7}) 
to describe the azimuthal mean overturning circulation forced by the adiabatic 
rearrangement of PV that occurs during the instability of hollow PV towers.

\section{Transverse circulation equation in the $\sigma$-coordinate}  

     Now consider the special case of the sigma coordinate, $\eta=\sigma=1-(p/p_{_S})$, 
where $p_{_S}(r,t)$ is the pressure at the earth's surface.  
In the sigma coordinate the pseudo-density becomes $m=p_{_S}/g$, and   
$p_{_S} \hat{\dot{\sigma}}=p_{_S} \dot{\sigma}-\partial p/\partial t$. 
The governing equations (\ref{eq2.1})--(\ref{eq2.5}) take the form    
\begin{equation}                           
        \left(f + \frac{v}{r}\right)v = \frac{\partial\Phi}{\partial r}
	                              + (1-\sigma)\alpha\frac{\partial p_{_S}}{\partial r},        
\label{eq8.1}
\end{equation}
\begin{equation}                           
                  \frac{\partial v}{\partial t} 
   + \dot{\sigma} \frac{\partial v}{\partial\sigma} 
   + \left(f + \frac{\partial(rv)}{r\,\partial r}\right)u = 0,     
\label{eq8.2}
\end{equation}
\begin{equation}                          
      \frac{\partial\Phi}{\partial\sigma} = p_{_S}\alpha,        
\label{eq8.3}
\end{equation}
\begin{equation}                          
          \frac{\partial p_{_S}}{\partial t}
        + \frac{\partial (p_{_S}ru)}{r\,\partial r}
        + \frac{\partial (p_{_S}\dot{\sigma})}{\partial\sigma} = 0,     
\label{eq8.4}
\end{equation}
\begin{equation}                          
                  \frac{\partial\theta}{\partial t} 
    +           u \frac{\partial\theta}{\partial r} 
    + \dot{\sigma}\frac{\partial\theta}{\partial\sigma} = \dot{\theta},  
\label{eq8.5}
\end{equation}
where in this section the partial derivatives $(\partial/\partial t)$ 
and $(\partial/\partial r)$ are taken at fixed $\sigma$.
The thermal wind equation (\ref{eq3.4}) takes the Jacobian form   
\begin{equation}                                  
        \hat{f}\frac{\partial v}{\partial\sigma} 
      = \frac{\partial(\Pi,\theta)}{\partial(r,\sigma)}.  
\label{eq8.6}
\end{equation}
The mass flux (\ref{eq3.3}) becomes 
\begin{equation}                                    
       (p_{_S}/g) u      = -\frac{\partial \psi}{\partial\sigma},  \qquad  
       (p_{_S}/g)\dot{\sigma} - \frac{1}{g}\frac{\partial p}{\partial t} = \frac{\partial(r\psi)}{r\,\partial r}.   
\label{eq8.7}
\end{equation}
The transverse circulation equation becomes  
\begin{equation}                                          
  \begin{split}
    & \frac{\partial}{\partial r}
       \left( A\frac{\partial(r\psi)}{r\,\partial r} 
            + B\frac{\partial \psi}{\partial\sigma}\right)     \\
    & \qquad 
     + \frac{\partial}{\partial\sigma}
       \left( B\frac{\partial(r\psi)}{r\,\partial r} 
            + C\frac{\partial  \psi }{ \partial\sigma}\right)   
     = \frac{\partial(\Pi,\dot{\theta})}{\partial(r,\sigma)},    
  \end{split}
\label{eq8.8}
\end{equation}
where, using the hydrostatic equation (\ref{eq8.3}), the coefficients 
(\ref{eq3.10}) become    
\begin{equation}                                         
  \begin{split}
     A &= \frac{g\alpha}{\theta}\frac{\partial\theta}{\partial\sigma},  \qquad 
     B  =-\frac{g\alpha}{\theta}\frac{\partial\theta}{\partial r},        \\
     C &= \frac{g\hat{f}}{p_{_S}}\left(f + \frac{\partial(rv)}{r\,\partial r}\right)
        - \frac{g}{p_{_S}}\frac{\partial\Pi}{\partial r}\frac{\partial\theta }{\partial r}.       
  \end{split}
\label{eq8.9}
\end{equation}
Note that $AC - B^2 = (g\alpha/\theta)\hat{f}P$, where
\begin{equation}                                        
      P = \frac{g}{p_{_S}}\left[-\frac{\partial v}{\partial\sigma}\frac{\partial\theta}{\partial r}    
        + \left(f + \frac{\partial(rv)}{r\,\partial r}\right)
	            \frac{\partial\theta}{\partial\sigma}\right].  
\label{eq8.10}
\end{equation}
An advantage of using $\sigma$ as the vertical coordinate  
is that the lower boundary condition for (\ref{eq8.8}) is applied on  
the coordinate surface $\sigma=0$. To the authors' knowledge the 
$\sigma$-coordinate form (\ref{eq8.8}) has not been used in studies 
of tropical cyclone dynamics.

\bigskip
\centerline{\bf Part II: Zonally Symmetric Meridional Circulations}
\medskip

    The purpose of Part II is to derive the meridional circulation equation 
for the zonally symmetric balanced model in the generalized vertical coordinate. 
Two important applications are the tropical Hadley circulation and the 
radiatively forced meridional circulation over Antarctica. 
We begin by reviewing the zonally symmetric balance equations 
in section 9. Section 10 presents a derivation of the meridional circulation 
equation, with (\ref{eq10.11}) being the main result. From (\ref{eq10.11}) 
the meridional circulation equations in $z$, $\hat{z}$, $\tilde{z}$, 
$\theta$, and $\sigma$ can be obtained as special cases.

\section{The zonally symmetric balanced model in $(\phi,\eta,t)$}  

     To simplify the primitive equation model to a balanced model 
we assume that the zonal flow remains in a nearly balanced state, 
i.e., we discard the meridional equation of motion and replace it 
with the balance condition given below in (\ref{eq9.2}).  A sufficient 
condition for the validity of this assumption is that the diabatic forcing 
term $\dot{\theta}$ has a slow enough time scale that significant, zonal  
mean inertia-gravity waves are not excited, i.e., $|Dv/Dt|$  
remains small compared to the pressure gradient and Coriolis/centrifugal 
terms. We shall describe this inviscid flow using a generalized vertical coordinate 
$\eta$, which is as yet an unspecified function of $(z,p,p_{_S},\theta)$. 
Under the balance condition, and using $\eta(z,p,p_{_S},\theta)$ as 
the vertical coordinate, the governing equations are  
\begin{equation}                           
                  \frac{\partial u}{\partial t} 
      + \dot{\eta}\frac{\partial u}{\partial\eta} 
      - \left(f - \frac{\partial(u\cos\phi)}{a\cos\phi\,\partial\phi}\right)v = 0,     
\label{eq9.1}
\end{equation}
\begin{equation}                           
        -\left(f + \frac{u\tan\phi}{a}\right)u 
             =       \frac{\partial\Phi}{a\partial\phi}
             + \theta\frac{\partial\Pi }{a\partial\phi},        
\label{eq9.2}
\end{equation}
\begin{equation}                          
      \frac{\partial\Phi}{\partial\eta} + \theta\frac{\partial\Pi}{\partial\eta} = 0,        
\label{eq9.3}
\end{equation}
\begin{equation}                          
          \frac{\partial m}{\partial t}
        + \frac{\partial (mv\cos\phi)}{a\cos\phi\,\partial\phi}
        + \frac{\partial (m \dot{\eta})}{\partial\eta} = 0,     
\label{eq9.4}
\end{equation}
\begin{equation}                          
                \frac{\partial\theta}{\partial t} 
    +         v \frac{\partial\theta}{a\partial\phi} 
    + \dot{\eta}\frac{\partial\theta}{\partial\eta} = \dot{\theta},  
\label{eq9.5}
\end{equation}
where $f=2\Omega\sin\phi$ is the Coriolis parameter, $m=-(1/g)(\partial p/\partial\eta)$ 
is the pseudodensity, $\alpha=RT/p$ the specific volume, $\Pi=c_p(p/p_0)^\kappa$ the 
Exner function, and where the partial derivatives $\partial/\partial t$ and 
$\partial/\partial\phi$ are understood to be at fixed $\eta$. In the following 
sections we consider only those cases for which the vertical coordinate $\eta$ 
increases in the upward direction, so that $m=-(1/g)(\partial p/\partial\eta) > 0$.  

\section{Meridional circulation equation in $(\phi,\eta,t)$}  

     To derive the meridional circulation equation associated with 
(\ref{eq9.1})--(\ref{eq9.5}), first consider the mass continuity equation 
(\ref{eq9.4}), which can be written in the form   
\begin{equation}                              
       \frac{\partial(mv\cos\phi)}{a\cos\phi\,\partial\phi} 
     + \frac{\partial(m\hat{\dot{\eta}})}{\partial\eta} = 0,  
\label{eq10.1}
\end{equation}
where 
\begin{equation}                                    
       m\hat{\dot{\eta}} = m\dot{\eta} - \frac{1}{g}\frac{\partial p}{\partial t}.  
\label{eq10.2}
\end{equation}
Using this form of the mass conservation principle we define a streamfunction 
$\psi$ such that 
\begin{equation}                                    
              m v       = -\frac{\partial \psi}{\partial\eta},  \qquad  
      m\hat{\dot{\eta}} =  \frac{\partial(\psi\cos\phi)}{a\cos\phi\,\partial\phi}.   
\label{eq10.3}
\end{equation}

     The thermal wind equation, derived from (\ref{eq9.2}) and (\ref{eq9.3}), is 
\begin{equation}                                  
        \hat{f}\frac{\partial u}{\partial\eta} 
      + \frac{\partial(\Pi,\theta)}{a\partial(\phi,\eta)} = 0,  
\label{eq10.4}
\end{equation}
where $\hat{f}=f+2(u/a)\tan\phi$. Taking $\partial/\partial t$ of (\ref{eq10.4}) 
we obtain 
\begin{equation}                                 
      \frac{\partial}{\partial\eta}\left(\hat{f}\frac{\partial u}{\partial t}\right) 
    + \frac{\partial}{\partial t}
        \left(\frac{\partial(\Pi,\theta)}{a\partial(\phi,\eta)}\right) = 0,   
\label{eq10.5}
\end{equation}
where we have made use of the identity 
$(\partial/\partial t)  [\hat{f}(\partial u/\partial\eta)]
=(\partial/\partial\eta)[\hat{f}(\partial u/\partial t  )]$ to express the 
left hand side in a form that has $(\partial u/\partial t)$ inside the 
$(\partial/\partial\eta)$ operator. This sets the stage for the elimination 
of $(\partial u/\partial t)$ through the use of (\ref{eq9.1}). Similarly, it 
is convenient to rewrite the second term of (\ref{eq10.5}) in a form that 
has local time derivatives inside the $(\partial/a\partial\phi)$ and 
$(\partial/\partial\eta)$ operators. This can be accomplished through the 
use of the Jacobi identity  
\begin{equation}                                     
    \frac{\partial}{a\partial\phi}\left(\frac{\partial(\Pi,\theta)}{ \partial(\eta, t  )}\right)
  + \frac{\partial}{ \partial\eta}\left(\frac{\partial(\Pi,\theta)}{a\partial(  t ,\phi)}\right)
  + \frac{\partial}{ \partial t  }\left(\frac{\partial(\Pi,\theta)}{a\partial(\phi,\eta)}\right) = 0,   
\label{eq10.6}
\end{equation}
which allows (\ref{eq10.5}) to be written in the form 
\begin{equation}                                     
    \frac{\partial}{a\partial\phi}\left(\frac{\partial(\Pi,\theta)}{ \partial(t,\eta)}\right)
  + \frac{\partial}{ \partial\eta}\left(\hat{f}\frac{\partial u}{\partial t}
                                      + \frac{\partial(\Pi,\theta)}{a\partial(\phi,t)}\right) = 0.    
\label{eq10.7}
\end{equation}
This is the constraint that must be satisfied by the local time derivatives of the 
zonal wind field and the mass field. This constraint will lead directly to the partial 
differential equation for $\psi$, after we have derived equations for the two quantities 
in the large parentheses of (\ref{eq10.7}). The equation for the first of these is derived 
by multiplying (\ref{eq9.5}) by $(\partial\Pi/\partial\eta)$, while the equation for 
the second is derived by adding $\hat{f}$ times (\ref{eq9.1}) 
to $(\partial\Pi/a\partial\phi)$ times (\ref{eq9.5}). The results are 
\begin{equation}                                   
         \frac{\partial(\Pi,\theta)}{\partial(t,\eta)}
     = -Am\hat{\dot{\eta}} + Bmv - \frac{\partial\Pi}{\partial\eta}\dot{\theta},   
\label{eq10.8}
\end{equation}
\begin{equation}                                   
     \hat{f}\frac{\partial u}{\partial t}
    + \frac{\partial(\Pi,\theta)}{a\partial(\phi,t)}
    = -Bm\hat{\dot{\eta}} + Cmv + \frac{\partial\Pi}{a\partial\phi}\dot{\theta},    
\label{eq10.9}
\end{equation}
where  
\begin{equation}                                   
  \begin{split}
     A &=  \frac{g\alpha}{\theta}\frac{\partial\theta}{\partial\eta},    \qquad 
     B  = -\frac{g\alpha}{\theta}\frac{\partial\theta}{a\partial\phi},   \\
     C &=  \frac{\hat{f}}{m}\left(f - \frac{\partial(u\cos\phi)}{a\cos\phi\,\partial\phi}\right) 
        - \frac{1}{m}\frac{\partial\Pi}{a\partial\phi}\frac{\partial\theta}{a\partial\phi}.  
  \end{split}
\label{eq10.10}
\end{equation}
Using (\ref{eq10.8}) and (\ref{eq10.9}) in (\ref{eq10.7}) we obtain the meridional 
circulation equation 
\begin{equation}                                           
 \begin{split}
    &  \frac{\partial}{a\partial\phi}
         \left( A \frac{\partial(\psi\cos\phi)}{a\cos\phi\,\partial\phi} 
              + B \frac{\partial \psi}{\partial\eta}\right)    \\
    & \quad 
     + \frac{\partial}{\partial\eta}
         \left( B \frac{\partial(\psi\cos\phi)}{a\cos\phi\,\partial\phi} 
              + C \frac{\partial  \psi}{\partial\eta}\right)      
     = \frac{\partial(\Pi,\dot{\theta})}{a\partial(\phi,\eta)}. 
 \end{split}   
\label{eq10.11}
\end{equation}
Note that $ AC-B^2=(g\alpha/\theta)\hat{f}P$, where 
\begin{equation}                                         
      P = \frac{1}{m}\left[\frac{\partial u}{\partial\eta}\frac{\partial\theta}{a\partial\phi}    
        + \left(f - \frac{\partial(u\cos\phi)}{a\cos\phi\partial\phi}\right)\frac{\partial\theta}{\partial\eta}\right]    
\label{eq10.12}     
\end{equation}
is the potential vorticity. Thus, the ellipticity condition $AC-B^2>0$ is equivalent 
to the condition $\hat{f}P>0$. This condition is generally satisfied because both $\hat{f}$ 
and $P$ tend to be positive in the northern hemisphere, while both tend to be negative in 
the southern hemisphere. However, the condition can often be violated near the equator. 
For example, when the ITCZ is north of the equator, lower tropospheric air with negative 
$P$ is drawn across the equator into the northern hemisphere, while upper tropospheric air 
with positive $P$ is forced across the equator into the southern hemisphere. This tends 
to result in small regions where $\hat{f}P<0$, the regions being just north of the equator 
in the lower troposphere and just south of the equator in the upper troposphere. Some of 
the possible dynamical consequences of this non-ellipticity have been discussed by 
\citet{tomas97}. However, in the context of solving the meridional circulation equation 
(\ref{eq10.11}), experience has shown that iterative methods generally work well even when 
there are small non-elliptic regions in the domain. 

     In the next five sections we consider the special cases of the height coordinate  
($\eta=z$), the log-pressure coordinate ($\eta=\hat{z}$), the pseudo-height coordinate 
($\eta=\tilde{z}$), the isentropic coordinate ($\eta=\theta$), and the sigma coordinate 
($\eta=1-(p/p_{_S})$).

\section{The meridional circulation equation in the height coordinate}  

     First consider the special case of the height coordinate, $\eta=z$. 
In the height coordinate the pseudo-density equals the true density, i.e., 
$m=\rho$.  The governing equations (\ref{eq9.1})--(\ref{eq9.5}) become   
\begin{equation}                           
         \frac{\partial u}{\partial t} 
      + w\frac{\partial u}{\partial z} 
      - \left(f - \frac{\partial(u\cos\phi)}{a\cos\phi\,\partial\phi}\right)v = 0,     
\label{eq11.1}
\end{equation}
\begin{equation}                           
        -\left(f + \frac{u\tan\phi}{a}\right)u 
             = \theta\frac{\partial\Pi}{a\partial\phi},        
\label{eq11.2}
\end{equation}
\begin{equation}                          
      g + \theta\frac{\partial\Pi}{\partial z} = 0,        
\label{eq11.3}
\end{equation}
\begin{equation}                          
          \frac{\partial\rho}{\partial t}
        + \frac{\partial (\rho v\cos\phi)}{a\cos\phi\,\partial\phi}
        + \frac{\partial (\rho w)}{\partial z} = 0,     
\label{eq11.4}
\end{equation}
\begin{equation}                          
        \frac{\partial\theta}{\partial t} 
    + v \frac{\partial\theta}{a\partial\phi} 
    + w \frac{\partial\theta}{\partial z} = \dot{\theta},   
\label{eq11.5}
\end{equation}
and the thermal wind equation (\ref{eq10.4}) becomes  
\begin{equation}                                  
        \hat{f}\frac{\partial u}{\partial z} 
      + \frac{\partial(\Pi,\theta)}{a\partial(\phi,z)} = 0.   
\label{eq11.6}
\end{equation}
The meridional circulation equation (\ref{eq10.11}) becomes  
\begin{equation}                              
  \begin{split}
    & \frac{\partial}{a\partial\phi}
       \left( A \frac{\partial(\psi\cos\phi)}{a\cos\phi\,\partial\phi} 
            + B \frac{\partial  \psi}{\partial z}\right)     \\
    & \qquad 
    + \frac{\partial}{\partial z}
       \left( B \frac{\partial(\psi\cos\phi)}{a\cos\phi\,\partial\phi} 
            + C \frac{\partial \psi}{\partial z}\right)
    = \frac{\partial(\Pi,\dot{\theta})}{a\partial(\phi,z)},    
  \end{split}
\label{eq11.7}
\end{equation}
where  
\begin{equation}                              
  \begin{split}
     \rho A &= \frac{g}{\theta}\frac{\partial\theta}{\partial z},   \qquad 
     \rho B  =-\frac{g}{\theta}\frac{\partial\theta}{a\partial\phi},        \\
     \rho C &= \hat{f}\left(f - \frac{\partial(u\cos\phi)}{a\cos\phi\,\partial\phi}\right)
            - \frac{\partial\Pi   }{a\partial\phi}
              \frac{\partial\theta}{a\partial\phi}.    
  \end{split}
\label{eq11.8}
\end{equation}
The mass flux (\ref{eq10.3}) becomes  
\begin{equation}                             
      \rho v =-\frac{\partial \psi}{\partial z},  \qquad  
      \rho w - \frac{1}{g}\frac{\partial p}{\partial t} 
             = \frac{\partial(\psi\cos\phi)}{a\cos\phi\,\partial\phi},   
\label{eq11.9}
\end{equation}
and that $ AC-B^2=(g\alpha/\theta)\hat{f}P$, where 
\begin{equation}                                   
    P = \frac{1}{\rho}\left[\frac{\partial u}{\partial z}\frac{\partial\theta}{a\partial\phi}    
      + \left(f-\frac{\partial(u\cos\phi)}{a\cos\phi\,\partial\phi}\right)\frac{\partial\theta}{\partial z}\right]. 
\label{eq11.10}
\end{equation} 
We are not aware of any studies of meridional circulation using the $z$-coordinate 
formulation (\ref{eq11.7}).

\section{The meridional circulation equation in the log-pressure coordinate}  
 
    Now consider the special case of the log-pressure coordinate, 
$\eta=\hat{z}=H\ln(p_0/p)$, where as before the scale height $H=RT_0/g$, the 
reference temperature $T_0$, and the reference pressure $p_0$ are constants.  
In this coordinate the pseudodensity becomes 
$m=\hat{\rho}(\hat{z})=\rho_0 e^{-\hat{z}/H}$, where 
the constant reference density is $\rho_0=p_0/RT_0$.  
The governing equations (\ref{eq9.1})--(\ref{eq9.5}) become   
\begin{equation}                           
               \frac{\partial u}{\partial t} 
      + \hat{w}\frac{\partial u}{\partial\hat{z}} 
      - \left(f - \frac{\partial(u\cos\phi)}{a\cos\phi\,\partial\phi}\right)v = 0,     
\label{eq12.1}
\end{equation}
\begin{equation}                           
     -\left(f + \frac{u\tan\phi}{a}\right)u = \frac{\partial\Phi}{a\partial\phi},        
\label{eq12.2}
\end{equation}
\begin{equation}                          
      \frac{\partial\Phi}{\partial\hat{z}} = \frac{g}{T_0}T,        
\label{eq12.3}
\end{equation}
\begin{equation}                          
       \frac{\partial(\hat{\rho} v\cos\phi)}{a\cos\phi\,\partial\phi}
     + \frac{\partial(\hat{\rho} w)}{\partial\hat{z}} = 0,     
\label{eq12.4}
\end{equation}
\begin{equation}                          
              \frac{\partial\theta}{\partial t} 
    +       v \frac{\partial\theta}{a\partial\phi} 
    + \hat{w} \frac{\partial\theta}{\partial\hat{z}} = \dot{\theta},  
\label{eq12.5}
\end{equation}
and the thermal wind equation (\ref{eq10.4}) loses its Jacobian form to become  
\begin{equation}                                  
        \hat{f}\frac{\partial u}{\partial\hat{z}} 
      = -\frac{g}{T_0}\frac{\partial T}{a\partial\phi}.  
\label{eq12.6}
\end{equation}
The mass flux (\ref{eq10.3}) becomes 
\begin{equation}                                    
             \hat{\rho} v = -\frac{\partial \psi}{\partial\hat{z}},  \qquad  
        \hat{\rho}\hat{w} =  \frac{\partial(\psi\cos\phi)}{a\cos\phi\,\partial\phi}.   
\label{eq12.7}
\end{equation}
The last term in the second line of (\ref{eq10.10}) vanishes 
and the transverse circulation equation (\ref{eq10.11}) becomes  
\begin{equation}                                       
  \begin{split}
    & \frac{\partial}{a\partial\phi}
       \left( A \frac{\partial(\psi\cos\phi)}{a\cos\phi\,\partial\phi} 
            + B \frac{\partial  \psi }{ \partial\hat{z}}\right)  \\
    & \qquad 
     + \frac{\partial}{\partial\hat{z}}
       \left( B \frac{\partial(\psi\cos\phi)}{a\cos\phi\,\partial\phi} 
            + C \frac{\partial  \psi }{\partial\hat{z}}\right)
    = \frac{T}{\theta}\frac{g}{T_0}\frac{\partial\dot{\theta}}{a\partial\phi},    
  \end{split}
\label{eq12.8}
\end{equation}
where 
\begin{equation}                                      
  \begin{split}
     \hat{\rho}A &= \frac{g}{T_0}\frac{\partial\theta}{\partial\hat{z}},   \quad 
     \hat{\rho}B  =-\frac{g}{T_0}\frac{\partial T}{a\partial\phi}
                  = \hat{f} \frac{\partial u}{\partial\hat{z}},  \\ 
     \hat{\rho}C &= \hat{f} \left(f - \frac{\partial(u\cos\phi)}{a\cos\phi\,\partial\phi}\right).   
  \end{split}
\label{eq12.9}
\end{equation}
Note that $AC - B^2 = (g\alpha/\theta)\hat{f}P$, where
\begin{equation}                                        
      P = \frac{1}{\hat{\rho}}
          \left[\frac{\partial u}{\partial\hat{z}}\frac{\partial\theta}{a\partial\phi}    
                + \left(f - \frac{\partial(u\cos\phi)}{a\cos\phi\,\partial\phi}\right)
	                    \frac{\partial\theta}{\partial\hat{z}}\right]    
\label{eq12.10}
\end{equation}
is the potential vorticity. A disadvantage of using $\hat{z}$ as the vertical coordinate 
is that the lower boundary condition for (\ref{eq12.8}) is not applied on a coordinate 
surface. Recent studies of the deep and shallow tropical Hadley circulations using a 
simplified, equatorial $\beta$-plane version of (\ref{eq12.8}) are those of \citet{gonzalez14} 
and \citet{gonzalez17}.

\section{The meridional circulation equation in the pseudo-height coordinate}  

     As in section 6, now consider the special case of the pseudo-height coordinate, 
$\eta=\tilde{z}$, where 
$\tilde{z} = (c_p\theta_0/g) \left[1 - \left(p/p_0\right)^\kappa\right]$. 	    
As before, in this coordinate the pseudo-density is given by 
\begin{equation}                                
    m = \tilde{\rho}(\tilde{z}) = \frac{p_0}{R\theta_0}
                              \left(1 - \frac{g\tilde{z}}{c_p \theta_0}\right)^{(1-\kappa)/\kappa}.     
\label{eq13.1}
\end{equation}
The governing equations (\ref{eq9.1})--(\ref{eq9.5}) become   
\begin{equation}                           
                 \frac{\partial u}{\partial t} 
      + \tilde{w}\frac{\partial u}{\partial\tilde{z}} 
      - \left(f - \frac{\partial(u\cos\phi)}{a\cos\phi\,\partial\phi}\right)v = 0,     
\label{eq13.3}
\end{equation}
\begin{equation}                           
     -\left(f + \frac{u\tan\phi}{a}\right)u = \frac{\partial\Phi}{a\partial\phi},        
\label{eq13.4}
\end{equation}
\begin{equation}                          
      \frac{\partial\Phi}{\partial\tilde{z}} = \frac{g}{\theta_0}\theta,        
\label{eq13.5}
\end{equation}
\begin{equation}                          
       \frac{\partial (\tilde{\rho} v\cos\phi)}{a\cos\phi\,\partial\phi}
     + \frac{\partial (\tilde{\rho} w)}{\partial\tilde{z}} = 0,     
\label{eq13.6}
\end{equation}
\begin{equation}                          
                \frac{\partial\theta}{\partial t} 
    +         v \frac{\partial\theta}{a\partial\phi} 
    + \tilde{w} \frac{\partial\theta}{\partial\tilde{z}} = \dot{\theta}.   
\label{eq13.7}
\end{equation}

    The meridional circulation equation (\ref{eq10.11}) becomes  
\begin{equation}                                 
  \begin{split}
    & \frac{\partial}{a\partial\phi}
       \left( A \frac{\partial(\psi\cos\phi)}{a\cos\phi\,\partial\phi} 
            + B \frac{\partial  \psi}{\partial\tilde{z}}\right)     \\
    & \quad 
    + \frac{\partial}{\partial\tilde{z}}
       \left( B \frac{\partial(\psi\cos\phi)}{a\cos\phi\,\partial\phi} 
            + C \frac{\partial  \psi}{\partial\tilde{z}}\right)
    = \frac{g}{\theta_0}\frac{\partial\dot{\theta}}{a\partial\phi},    
  \end{split}
\label{eq13.8}
\end{equation}
where  
\begin{equation}                              
  \begin{split}  
    \tilde{\rho}A &= \frac{g}{\theta_0}\frac{\partial\theta}{\partial\tilde{z}},   \qquad 
    \tilde{\rho}B  =-\frac{g}{\theta_0}\frac{\partial\theta}{a\partial\phi}
                   = \hat{f}\frac{\partial u}{\partial\tilde{z}},        \\
    \tilde{\rho}C &= \hat{f}\left(f - \frac{\partial(u\cos\phi)}{a\cos\phi\,\partial\phi}\right).        
  \end{split}
\label{eq13.9}
\end{equation}
The mass flux (\ref{eq10.3}) becomes
\begin{equation}                             
      \tilde{\rho}  v       = -\frac{\partial \psi}{\partial\tilde{z}},  \qquad  
      \tilde{\rho}\tilde{w} =  \frac{\partial(\psi\cos\phi)}{a\cos\phi\,\partial\phi}, 
\label{eq13.10}
\end{equation}
and $ AC-B^2=(\alpha/\theta)\hat{f}P$, where 
\begin{equation}                             
    P = \frac{1}{\tilde{\rho}}\left[\frac{\partial u}{\partial\tilde{z}}
                                    \frac{\partial\theta}{a\partial\phi}    
      + \left(f -\frac{\partial(u\cos\phi)}{a\cos\phi\,\partial\phi}\right)
                 \frac{\partial\theta}{\partial\tilde{z}}\right].   
\label{eq13.11}
\end{equation}  
\citet{hack+schubert+stevens+kuo89} and \citet{hack+schubert90} have solved 
(\ref{eq13.8}) for cases in which the diabatic forcing $\dot{\theta}$ is associated 
with an ITCZ that lies off the equator. This produces two Hadley cells, with the 
cross-equatorial cell carrying considerably more mass flux than its companion. 
This anisotropy is due to the spatial variation of the inertial stability coefficient 
$C$, which is relatively small near the equator, thereby providing little resistance 
to horizontal flow across the equator.

\section{The meridional circulation equation in $(\phi,\theta,t)$}  

     Now consider the special case of the isentropic coordinate, $\eta=\theta$. In 
the isentropic coordinate the pseudodensity becomes $m=-(1/g)(\partial p/\partial\theta)$ 
and $A=g\alpha/\theta$, $B=0$, and the last term in the second line of (\ref{eq10.10}) 
vanishes so that $C=\hat{f}P$. The governing equations (\ref{eq9.1})--(\ref{eq9.5}) become   
\begin{equation}                           
                    \frac{\partial u}{\partial t} 
      + \dot{\theta}\frac{\partial u}{\partial\theta} 
      - \left(f - \frac{\partial(u\cos\phi)}{a\cos\phi\,\partial\phi}\right)v = 0,     
\label{eq14.1}
\end{equation}
\begin{equation}                           
      -\left(f + \frac{u\tan\phi}{a}\right)u = \frac{\partial M}{a\partial\phi},        
\label{eq14.2}
\end{equation}
\begin{equation}                          
        \frac{\partial M}{\partial\theta} = \Pi,        
\label{eq14.3}
\end{equation}
\begin{equation}                          
          \frac{\partial m}{\partial t}
        + \frac{\partial (mv\cos\phi)}{a\cos\phi\,\partial\phi}
        + \frac{\partial (m\dot{\theta})}{\partial\theta} = 0,       
\label{eq14.4}
\end{equation}
and the thermal wind equation (\ref{eq10.4}) has the non-Jacobian form  
\begin{equation}                          
        \hat{f}\frac{\partial u}{\partial\theta} = \frac{\partial\Pi}{a\partial\phi}.         
\label{eq14.5}
\end{equation}
The meridional circulation equation (\ref{eq10.11}) becomes  
\begin{equation}                                         
      \frac{\partial}{a\partial\phi}
       \left( \frac{g\alpha}{\theta} \frac{\partial(\psi\cos\phi)}{a\cos\phi\,\partial\phi}\right) 
    + \frac{\partial}{\partial\theta}
       \left( \hat{f} P \frac{\partial\psi}{\partial\theta}\right)
    = \frac{\partial(\Pi,\dot{\theta})}{a\partial(\phi,\theta)},       
\label{eq14.5}
\end{equation}
where  
\begin{equation}                                         
      P = \left(f - \frac{\partial(u\cos\phi)}{a\cos\phi\,\partial\phi}\right)
          \left(-\frac{1}{g}\frac{\partial p}{\partial\theta}\right)^{-1}   
\label{eq14.6}
\end{equation}
is the potential vorticity. The mass flux (\ref{eq10.3}) becomes  
\begin{equation}                                         
  \begin{split}		      
   -(1/g)(\partial p/\partial\theta) v                 &= -\frac{\partial \psi}{\partial\theta},  \\   
   -(1/g)(\partial p/\partial\theta)\hat{\dot{\theta}} 
          - \frac{1}{g}\frac{\partial p}{\partial t}   &=  \frac{\partial(\psi\cos\phi)}{a\cos\phi\,\partial\phi}. 
  \end{split}
\label{eq14.7}
\end{equation}

     The advantages of using $\theta$ as the vertical coordinate for 
stratospheric studies have been discussed by \citet{andrews83}, \citet{tung86}, 
and \citet{yang+tung90}. Although (\ref{eq14.5}) and (\ref{eq14.6}) have 
compact forms, a disadvantage of using $\theta$ as the vertical coordinate 
for tropospheric studies is that the lower boundary condition for (\ref{eq14.5}) 
is not applied on a coordinate surface. As discussed in section 7, this 
disadvantage is sometimes overcome through the use of a massless layer.  
\citet{fulton17} have applied the concept of a massless layer in solving the 
PV invertibility principle for the topographically bound low-level jet surrounding 
Antarctica.

\section{The meridional circulation equation in $(\phi,\sigma,t)$}    

     Now consider the special case of the sigma coordinate, $\eta=\sigma=1-(p/p_{_S})$, 
where $p_{_S}(\phi,t)$ is the pressure at the earth's surface.  
In the sigma coordinate the pseudodensity becomes $m=p_{_S}/g$, and   
$p_{_S} \hat{\dot{\sigma}}=p_{_S} \dot{\sigma}-\partial p/\partial t$. 
The governing equations (\ref{eq9.1})--(\ref{eq9.5}) become   
\begin{equation}                           
                    \frac{\partial u}{\partial t} 
      + \dot{\sigma}\frac{\partial u}{\partial\sigma} 
      - \left(f - \frac{\partial(u\cos\phi)}{a\cos\phi\,\partial\phi}\right)v = 0,     
\label{eq15.1}
\end{equation}
\begin{equation}                           
        -\left(f + \frac{u\tan\phi}{a}\right)u 
       = \frac{\partial\Phi}{a\partial\phi} + \theta\frac{\partial\Pi}{a\partial\phi},        
\label{eq15.2}
\end{equation}
\begin{equation}                          
      \frac{\partial\Phi}{\partial\sigma} + \theta\frac{\partial\Phi}{\partial\sigma} = 0,        
\label{eq15.3}
\end{equation}
\begin{equation}                          
          \frac{\partial m}{\partial t}
        + \frac{\partial (mv\cos\phi)}{a\cos\phi\,\partial\phi}
        + \frac{\partial (m\dot{\sigma})}{\partial\sigma} = 0,     
\label{eq15.4}
\end{equation}
\begin{equation}                          
                  \frac{\partial\theta}{\partial t} 
    +            v\frac{\partial\theta}{a\partial\phi} 
    + \dot{\sigma}\frac{\partial\theta}{\partial\sigma} = \dot{\theta},  
\label{eq15.5}
\end{equation}
and the thermal wind equation (\ref{eq10.4}) has the Jacobian form  
\begin{equation}                          
     \hat{f}\frac{\partial u}{\partial\sigma} = \frac{\partial(\Pi,\theta)}{a\partial(\phi,\sigma)}.        
\label{eq15.6}
\end{equation}
The meridional circulation equation (\ref{eq10.11}) becomes  
\begin{equation}                                          
  \begin{split}
    & \frac{\partial}{a\partial\phi}
       \left( A\frac{\partial(\psi\cos\phi)}{a\cos\phi\,\partial\phi} 
            + B\frac{\partial \psi}{\partial\sigma}\right)     \\
    & \qquad 
     + \frac{\partial}{\partial\sigma}
       \left( B\frac{\partial(\psi\cos\phi)}{a\cos\phi\,\partial\phi} 
            + C\frac{\partial  \psi }{ \partial\sigma}\right)   
     = \frac{\partial(\Pi,\dot{\theta})}{a\partial(\phi,\sigma)},    
  \end{split}
\label{eq15.7}
\end{equation}
where  
\begin{equation}                                         
  \begin{split}
    A &= \frac{g\alpha}{\theta}\frac{\partial\theta}{\partial\sigma},  \qquad 
    B  =-\frac{g\alpha}{\theta}\frac{\partial\theta}{a\partial\phi},        \\
    C &= \frac{g\hat{f}}{p_{_S}}
          \left(f - \frac{\partial(u\cos\phi)}{a\cos\phi\,\partial\phi}\right)
        - \frac{gp\alpha}{p_{_S}}\frac{\partial\ln p_{_B}}{a\partial\phi}
                                 \frac{\partial\ln\theta}{a\partial\phi}.       
  \end{split}
\label{eq15.8}
\end{equation}
The mass flux (\ref{eq10.3}) becomes
\begin{equation}                                         
      (p_{_S}/g) v                 = -\frac{\partial \psi}{\partial\sigma},  \qquad  
      (p_{_S}/g)\dot{\sigma} - \frac{1}{g}\frac{\partial p_{_S}}{\partial t} 
                                   =  \frac{\partial(\psi\cos\phi)}{a\cos\phi\,\partial\phi}, 
\label{eq15.9}
\end{equation}
and that $AC - B^2 = (g\alpha/\theta)\hat{f}P$, where
\begin{equation}                                        
      P = \frac{g}{p_{_S}}\left[\frac{\partial u}{\partial\sigma}\frac{\partial\theta}{a\partial\phi}    
        + \left(f - \frac{\partial(u\cos\phi)}{a\cos\phi\,\partial\phi}\right)
	            \frac{\partial\theta}{\partial\sigma}\right].  
\label{eq15.10}
\end{equation}
An advantage of using $\sigma$ as the vertical coordinate 
is that the lower boundary condition for (\ref{eq15.7}) is applied on  
the coordinate surface $\sigma=0$. Thus, an interesting application of 
(\ref{eq15.7}) would be to the katabatically forced meridional circulation 
over Antarctica.

\section{Concluding Remarks}                      

     Using the generalized vertical coordinate $\eta(z,p,p_{_S},\theta)$, we have 
derived the transverse circulation equation (\ref{eq3.11}) for tropical cyclones 
and the meridional circulation equation (\ref{eq10.11}) for zonally symmetric flows. 
These derivations involve the time derivative of the thermal wind equation, which 
provides a constraint on the tendencies of the rotational wind field and the mass 
field. A crucial mathematical step in these derivations is to obtain the constraints 
in the forms (\ref{eq3.7}) and (\ref{eq10.7}), i.e., with spatial derivatives on 
the outside and time derivatives on the inside. This relies on the use of the Jacobi 
identities (\ref{eq3.6}) and (\ref{eq10.6}), which can thus be considered the key 
step in producing the final partial differential equations (\ref{eq3.11}) and 
(\ref{eq10.11}). 

    The solutions of the elliptic problems discussed here depend on both the forcing 
and the boundary conditions. For simplicity we have not discussed frictional 
forcing and boundary conditions. In many applications, the elliptic equations we have 
derived would be applied to an inviscid interior flow, with a separate frictional 
boundary layer dynamics providing the lower boundary condition for the inviscid interior. 
The boundary layer dynamics could involve simple, local, slab Ekman theory over a flat 
surface for the tropical cyclone case or over a sloping ice surface for the Antarctic 
case. In any event, the formulation of the lower boundary condition is beyond the 
scope of the present analysis. For an interesting discussion of some lower boundary 
effects the reader is referred to the interesting study by \citet{haynes89}, who 
discussed the importance of surface pressure changes in the response of the 
atmosphere to zonally-symmetric thermal and mechanical forcing. 

     For simplicity we have also omitted discussion of hybrid vertical coordinates, such 
as those proposed by \citet{zhu92} and \citet{konor97}. However, since these hybrid 
coordinates are also special cases of the generalized vertical coordinate 
$\eta(z,p,p_{_S},\theta)$, transverse circulation equations can easily be obtained 
from the generalized forms (\ref{eq3.11}) and (\ref{eq10.11}).  


%


%

\begin{acknowledgment}
     We would like to thank Rick Taft and Christopher Slocum 
for their comments. This research has been supported by the National Science 
Foundation under a collaborative grant with award numbers AGS-1147431
and AGS-1601628 to Clarkson University and AGS-1147120 and AGS-1601623 to
Colorado State University. 

\end{acknowledgment}

\ifthenelse{\boolean{dc}}
{}
{\clearpage}
\bibliographystyle{ametsoc2014}
\bibliography{references}


%
\end{document}